\documentclass[twocolumn,aps,prl,showpacs,superscriptaddress,floatfix,nofootinbib]{revtex4-2}

\usepackage{algorithm}
\usepackage{algorithmic}
\usepackage{float}
\floatstyle{ruled}
\newfloat{algorithm}{htbp}{loa}
\floatname{algorithm}{Algorithm}

\usepackage[utf8]{inputenc} 
\usepackage[T1]{fontenc}    
\usepackage{hyperref}       
\usepackage{url}            
\usepackage{booktabs}       
\usepackage{amsfonts}       
\usepackage{nicefrac}       
\usepackage{microtype}      
\usepackage{xcolor}         
\usepackage{graphicx}
\usepackage{amsmath}
\usepackage[normalem]{ulem} 

\newcommand\sect[1]{\noindent \textit{\textbf{#1.}}---}

\newcommand\be{\begin{equation}}
\newcommand\ee{\end{equation}}

\date{\today}

\begin{document}

\title{An end-to-end generative diffusion model for heavy-ion collisions}

\author{Jing-An Sun}
\affiliation{Institute of Modern Physics, Fudan University, Handan Road 220, Yangpu District, Shanghai,
200433, China}
\affiliation{Department of Physics, McGill University, Montreal, Quebec H3A 2T8, Canada}
\author{Li Yan}
\affiliation{Institute of Modern Physics, Fudan University, Handan Road 220, Yangpu District, Shanghai,
200433, China}
\affiliation{Key Laboratory of Nuclear Physics and Ion-beam Application (MOE), Fudan University, Shanghai 200433, China}
\author{Charles Gale}
\affiliation{Department of Physics, McGill University, Montreal, Quebec H3A 2T8, Canada}
\author{Sangyong Jeon}
\affiliation{Department of Physics, McGill University, Montreal, Quebec H3A 2T8, Canada}
 
\begin{abstract}
We train a generative diffusion model (DM) to simulate ultra-relativistic heavy-ion collisions from end to end. The model takes initial entropy density profiles as input and produces two-dimensional final particle spectra, successfully reproducing integrated and differential observables. 
It also captures higher-order fluctuations and correlations. 
These findings suggest that the generative model has successfully learned the complex relationship between initial conditions and final particle spectra for various shear viscosities, as well as the fluctuations introduced during initial entropy production and hadronization stages, providing an efficient framework for resource-intensive physical goals. The code and trained model are available at \href{[https://huggingface.co/Jing-An/DiffHIC/tree/main}{https://huggingface.co/Jing-An/DiffHIC/tree/main}.

\end{abstract}

\maketitle

\sect{Introduction}
The high energy heavy-ion collisions carried out at the Large Hadron Collider (LHC) and the Relativistic Heavy Ion Collider (RHIC) create a new state of matter, the quark-gluon plasma (QGP)~\cite{Shuryak:2014zxa,Heinz:2013th}. QGP is fluid-like, which makes the theoretical modeling based on hydrodynamics~\cite{hybridBass:2000ib,hybridHirano:2007ei,hybridNonaka:2006yn,hybridPetersen:2008dd,hybridRyu:2017qzn,hybridSong:2010aq,hybridZhu:2015dfa,Yan:2017ivm,Shen:2020mgh} remarkably successful. Most contemporary studies implement a so-called hybrid approach, where event-by-event an initial entropy distribution~\cite{ICVredevoogd:2008id,ICmueller2011entropy,ICSchenke:2012wb,ICvanderSchee:2013pia,ICBerges:2014yta,ICKurkela:2014tea,ICKurkela:2018vqr,ICKurkela:2019set,ICKurkela:2018wud,ICSchlichting:2019abc} is followed by viscous relativistic hydrodynamic evolution~\cite{hydroSchenke:2010rr,hydroSchenke:2010nt,hydroPaquet:2015lta,hydroGale:2021emg,Romatschke:2009im,Florkowski:2017olj,Gale:2013da}, which dovetails into relativistic hadronic transport~\cite{SMASH:2016zqf,urqmdBass:1998ca,urqmdBleicher:1999xi,urqmdPetersen:2008dd}. In such models, particle spectra from experiments can be well-described, and so can the various signatures of collective flow, flow correlations, and fluctuations~\cite{Ollitrault:1992bk}.  

Despite this success, the traditional numerical simulations of hydrodynamics struggle to confront recent high-precision measurements. In experiments,  the data from $10^9\sim 10^{10}$ collision events~\cite{whiteArslandok:2023utm,reviewALICE:2022wpn} allow one to probe the finer details in the system, such as the nuclear structure
~\cite{pTFortier:2023xxy,pTv2Jia:2021wbq,pTv2Bally:2021qys,pTv2Bozek:2016yoj} and speed of sound in QGP~\cite{cs2Gardim:2019brr,cs2Sun:2024zsy,cs2CMS:2024sgx,cs2Nijs:2023bzv}, via statistics-demanding observables. It is quite challenging for theoretical model calculations to achieve comparable precision, as the traditional numerical simulation of hydrodynamics for one central event typically takes approximately 120 minutes ($10^4$ seconds) on a single CPU. As heavy-ion collision physics enters a high-precision era, theoretical modeling needs to evolve to meet the growing computational demands.



As a natural solution, machine learning (ML) and artificial intelligence (AI) have emerged as promising tools to optimize and enhance real-time predictions in hydrodynamic simulations for heavy-ion collisions~\cite{Heffernan:2023gye,fastHuang:2018fzn,fastLiyanage:2022byj}. Specifically, we propose using diffusion models (DMs)~\cite{Diffho2020denoising,Diffsohl2015deep,Diffsong2020score} as a powerful approach that excels in capturing the complex dynamics of heavy-ion systems, in order to detect the possibly intricate correlations between initial entropy density and particle spectra, while extracting transport properties of the QGP.

DMs~\cite{Diffho2020denoising,Diffsohl2015deep,Diffsong2020score}, a promising class of the generative models, demonstrate efficacy in mapping randomly sampled Gaussian noise to complex target distributions~\cite{BeatGANdhariwal2021diffusion,BeatGANho2022cascaded}. Compared to generative adversarial networks (GANs), diffusion models offer high-quality generation and excellent model convergence~\cite{Conversong2021maximum,Converhuang2021variational,Converkingma2021variational}. In heavy-ion physics, where particle spectra are multi-dimensional distributions, DMs are expected {to be}  particularly well-suited for event-by-event generation.
Note that DMs
have {also} seen applications in detector simulations for heavy-ion experiments~\cite{expAmram:2023onf,expGo:2024xor,expLeigh:2023toe,expLeigh:2023zle,expMikuni:2023dvk,expMikuni:2023tqg,expShmakov:2023kjj,mlHe:2023zin,mlMa:2023zfj,mlPang:2024kid,mlZhou:2023pti}.

In this Letter, we {introduce} DiffHIC (Diffusion Model for Heavy-Ion Collisions), a novel generative diffusion model developed to generate final state {two-dimensional} charged particle spectra, based on initial conditions and transport parameters. This marks the first application of a {diffusive} generative model to the simulation of heavy-ion collisions. By comparing observables derived from particle spectra generated by both traditional numerical simulations and our trained generative model, we demonstrate that DiffHIC not only accurately replicates integrated and differential observables but also effectively captures higher-order fluctuations and correlations. These results indicate that DiffHIC successfully learns the intricate mapping from initial entropy density profiles to final particle spectra, governed by a set of nonlinear hydrodynamic and Boltzmann transport equations. While preserving the intricate details of the underlying physical processes, DiffHIC significantly accelerates end-to-end heavy-ion collision simulations. For example, DiffHIC accomplishes one single central collision event in just $10^{-1}$ seconds on a GeForce GTX 4090 GPU.

\sect{The generative diffusion model}
The generative diffusion model {comprises} a forward process and a reverse process. In the forward process, the original data distribution is transformed into a known prior, by gradually injecting noise. Such a process is governed by a stochastic differential equation (SDE)~\cite{Diffsong2020score},
\begin{align}
    d\pmb x = {\bf f}(\pmb x,t) dt + g(t) d\pmb w,
    \label{eq:nosing}
\end{align}
and correspondingly, a reverse-time SDE~\cite{ANDERSON1982313},
\begin{align}
    d\pmb{x} = [{\bf f}(\pmb{x}, t) - g(t)^2\nabla_{\pmb{x}} \log p_t(\pmb{x})]dt + g(t)d\bar{\pmb{w}},
    \label{eq:denoising}
\end{align}
transforms the prior distribution back into the data distribution by gradually
removing the noise. Here, $\pmb w$ and $\bar{\pmb{w}}$ both represent the standard Wiener processes (Gaussian white noise), with ${\bf f}(\pmb x,t)$ the drift coefficient and $g(t)$ the diffusion coefficient of $\pmb x(t)$. 
In the generative diffusion model, the probability distribution $p_t(\pmb x)$, hence the
score function $\nabla_{\pmb{x}} \log p_t(\pmb{x})$, is generally unknown, which can be estimated by a neural network with parameter $\pmb \theta$ via minimizing the explicit score matching (ESM) loss $\mathcal L_t^{\text{ESM}} \equiv \mathbb{E}_{p(\pmb x_t)} ||\pmb s_{\pmb \theta}(\pmb x,t)-\nabla_{\pmb{x}} \log p_t(\pmb{x})||^2$. Once we have a trained $s_{\theta} (\pmb x,t)$, the trajectory from the prior distribution to the real data distribution can be determined following Eq.~\ref{eq:denoising}.

While the goal in DMs is to estimate the score function, instead of minimizing \(\mathcal{L}_t^{\text{ESM}} \),  it is more feasible to solve from the denoising score matching loss (DSM), $\mathcal L_t^{\text{DSM}}\equiv \mathbb{E}_{p(\pmb x_t,\pmb x_0)}||\pmb{s}_{\pmb{\theta}}(\pmb{x}, t) - \nabla_{\pmb{x}} \log p(\pmb{x}_t | \pmb{x}_0)||^2$, because the conditional probability $p(\pmb{x}_t | \pmb{x}_0)$ is solvable by construction. 
Noticing that ESM and DSM differs by a shift that is independent from the parameter ${\pmb \theta}$,  minimization 
\begin{align}
    \pmb{\theta}^* &= \text{argmin}_{\pmb \theta} \mathcal L_t^{\text{DSM}} = \text{argmin}_{\pmb \theta} \mathcal L_t^{\text{ESM}}\,,
\end{align}
equivalently determines the score function. The equivalence can be intuitively understood from the relation $p_t({\bf x})=\int d {\bf x}_0 p({\bf x}|{\bf x}_0)$, where the parameters do not contribute.
In this Letter, we employ the so-called variance-preserving SDE~\cite{Diffsong2020score,Diffho2020denoising,BeatGANdhariwal2021diffusion}, in which with ${\bf f}(\pmb x,t) = -\frac{1}{2} \beta(t) \pmb x$ and $g(t) = \sqrt{\beta(t)}$ the conditional probability distribution is analytically solvable. After discretization, the forward SDE can be viewed as a Markov chain,
\begin{align}
    \pmb x (t_i) \approx \pmb x(t_{i-1}) - \frac{1}{2}\beta(t_i) \Delta t  \pmb x(t_{i-1}) + \sqrt{\beta(t_i)\Delta t} \pmb \varepsilon(t_i),
\end{align}
where $t_i=i/N$, and $N$ the number of steps. With respect to the properties of Wiener process, 
$\sqrt{\Delta t} \pmb \varepsilon(t_i)$ follows a Gaussian noise of zero mean and width $\Delta t$,  namely, $ \mathcal N(0,\Delta t)$. Let $\beta_i=N\beta(t_i), \pmb x (t_i) = \pmb x_i,  \pmb \varepsilon(t_i) = \pmb \varepsilon_i$, one has $p(\pmb{x}_i | \pmb{x}_0) = \mathcal{N}(\sqrt{\bar{\alpha}_i} \pmb{x}_0, \sigma_i^2 \pmb{I})$,
where \(\alpha_i = 1 - \beta_i\), \(\bar{\alpha}_i = \prod_{j=0}^i \alpha_j\) and \(\sigma_i = \sqrt{1 - \bar{\alpha}_i}\). Therefore, 
$
    \pmb{\theta}^* = \text{argmin}_{\pmb \theta } \mathbb{E}_{p(\pmb x_{t_i},\pmb x_0)} ||\pmb{s}_{\pmb{\theta}}(\pmb{x}, t_i) + \frac{\pmb{x}_i - \sqrt{\bar{\alpha}_i} \pmb{x}_0}{\sigma_i^2}||^2
    =\text{argmin}_{\pmb \theta} \mathbb{E}_{p(\pmb x_{t_i},\pmb x_0)}||\pmb{\varepsilon}_{\pmb{\theta}}(\pmb{x}, t_i) - \pmb{\varepsilon}||^2,
$
where \(\pmb{x}_i = \sqrt{\bar{\alpha}_i} \pmb{x}_0 + \sigma_i \pmb{\varepsilon}\) and \(\pmb{\varepsilon} \sim \mathcal{N}(0, \pmb{I})\). For simplicity, we have defined the scaled score function  \(\pmb{\varepsilon}_{\pmb{\theta}} = -\sigma_i \pmb{s}_{\pmb{\theta}}\) as the optimization target, referred to as the noise prediction network.

With the trained noise prediction network $\pmb\varepsilon_{\pmb\theta}$, one can generate the sample from a prior standard normal distribution via the solution of reverse SDE. However, stochasticity is introduced in the SDE solution, which will induce unphysical fluctuations in final particle spectra. Accordingly, we consider the corresponding probability flow ordinary differential equations (ODE)~\cite{Diffsong2020score},
\begin{align}\label{eq:reverseODE}
    d\pmb{x} = \left( {\bf f}(\pmb{x}, t) - \frac{1}{2}g(t)^2 \nabla_{\pmb{x}} \log p_t(\pmb{x}) \right) dt,
\end{align}
which converts the probabilistic models to the deterministic models. 
Because the reserve ODE and the SDE share the same probability distribution (cf. the Supplemential Material~\cite{SDE_ODE}), the sampling can be achieved identically, in terms of, for instance, fidelity, precision and full variability of the data. 

Moreover,
fast sampling can be performed through the numerical methods of ODE~\cite{zheng2023dpm,lu2023dpmsolverfastsolverguided,lu2022dpmsolver}.

In many generative modeling tasks, it is often desirable to guide the generation process by incorporating additional information, denoted as $\pmb y$. This approach is known as conditional generative modeling, where the goal is to model the conditional distribution $p(\pmb x|\pmb y)$, representing the generation of samples $\pmb x$ {conditional} on $\pmb y$. In the context of ultra-relativistic heavy-ion collisions, our goal is to train a conditional generative diffusion model that takes these initial entropy density profiles and shear viscosity as conditions to generate the final particle spectra. Analogous to unconditional diffusion models, this can be achieved by replacing the noise prediction network $\pmb{\varepsilon}_{\pmb\theta}(\pmb x_t,t)$ with $\pmb{\varepsilon}_{\pmb\theta}(\pmb x_t,\pmb y, t)$ to incorporate the conditional information into the generative process.

\sect{The hybrid approach to heavy-ion collision modeling}
We 
{briefly summarize} 
the hybrid model for heavy-ion collisions. At the initial time 
$\tau_0$, the entropy production is calculated with the 
{\textsc{TrENTo} model}~\cite{trentoMoreland:2014oya}, where fluctuations in the positions of the nucleons and the contributed entropy in each nucleon-nucleon collision have been 
{taken into account.}
The system subsequently 
{undergoes} 
hydrodynamic evolution  which is realized by \href{https://github.com/MUSIC-fluid/MUSIC/}{MUSIC}~\cite{hydroSchenke:2010nt,hydroSchenke:2010rr,hydroPaquet:2015lta} with a lattice QCD equation of state. In this work, we focus on the mid-rapidity region where the dynamics can be approximated as effectively (2+1)-dimensional with longitudinal boost-invariance. The bulk viscosity effect is neglected and the ratio of shear viscosity over entropy density is set to be $\eta/s = 0.0, 0.1,$ and $0.2$.
When the local energy density drops to a switching value $\varepsilon_{sw} = 0.18 \text{ GeV/fm}^3$, the transition from fluid 
to particles occurs through the Cooper-Frye formula~\cite{CF_Cooper:1974mv,CF_Huovinen:2012is}.
The particles with well-defined positions and momenta are 
{randomly} sampled from each fluid cell individually
{by using the publicly available iSS sampler} {\href{https://github.com/chunshen1987/iSS}{iSS}}. After particlization,  {\href{https://github.com/jbernhard/urqmd-afterburner}{UrQMD}} simulates the Boltzmann transport of all hadrons in the system and considers the rescatterings among hadrons and their excited resonance states, as well as all strong decay processes.

The model generates a dataset which covers the transverse momentum $p_T \in [0.0, 2.0]$ GeV and the rapidity $y \in [-0.5, 0.5]$. These ranges are chosen to align with the kinematic regions commonly studied in heavy-ion collision experiments, ensuring that the model captures the essential features of particle production.

\sect{A generative diffusion model for  heavy-ion collisions}
In this work, we train a generative diffusion model to function as a heavy-ion collision event generator. We carried out (2+1)D minimum bias simulations of Pb-Pb collisions at 5.02 TeV, choosing the shear viscosity \(\eta/s\) to be one of three distinct values: 0.0, 0.1, and 0.2. For each value of \(\eta/s\), we generate 12\,000 pairs of initial entropy density profiles and final particle spectra, corresponding to 12\,000 simulated events
{across all centralities}, 
as the training dataset. {The detailed parameters for the initial stage, hydrodynamic evolution, and 
the particlization are provided in the Supplemental Material~\cite{ModelParameters}.} As is a standard practice in machine learning, 70\% of the total events are used for training and the rest are used for validation. 

Considering that the spectra $\pmb S_0$ depend on the initial entropy density profiles $\pmb I$ and the shear viscosity $\eta/s$, we train a conditional reverse diffusion process $p(\pmb S_0|\pmb I,\eta/s)$ without modifying the forward process. 
\begin{algorithm}[H]
\caption{Training  DiffHIC}
\begin{algorithmic}
\STATE \textbf{Input:} Initial entropy profiles $\pmb I$, final particle spectra $\pmb S$ pairs, and corresponding shear viscosity $\eta/s$, number of diffusion steps $T$, noise schedule $\beta_t$
\STATE \textbf{Repeat}
\FOR{each training iteration}
    \STATE \hspace{1em} Sample pairs $(\pmb I,\pmb S_0,\eta/s)$ from the true data 
    \STATE \hspace{1em} Sample $t \sim \text{Uniform}(\{1, \dots, T\})$
    \STATE \hspace{1em} Sample $\pmb \varepsilon $ from standard normal distribution
    \STATE \hspace{1em} Compute noisy spectra $\pmb {S}_t = \sqrt{\bar \alpha_t} \pmb{S}_0 + \sqrt{1 - \bar \alpha_t} \pmb\varepsilon$
    \STATE \hspace{1em} Compute loss $\mathcal{L}_t({\pmb\theta}) = \|{\pmb\varepsilon} - 
    {\pmb\varepsilon}_{\pmb\theta}(\pmb {S}_t,\pmb I, t, \eta/s)\|^2$
    \STATE \hspace{1em} Update model parameters $\theta$ using gradient descent on $\mathcal{L}_t(\theta)$
\ENDFOR
\STATE \textbf{Until convergence}
\end{algorithmic}
\end{algorithm}

\begin{figure}[t]
    \centering
    \includegraphics[width=1\linewidth]{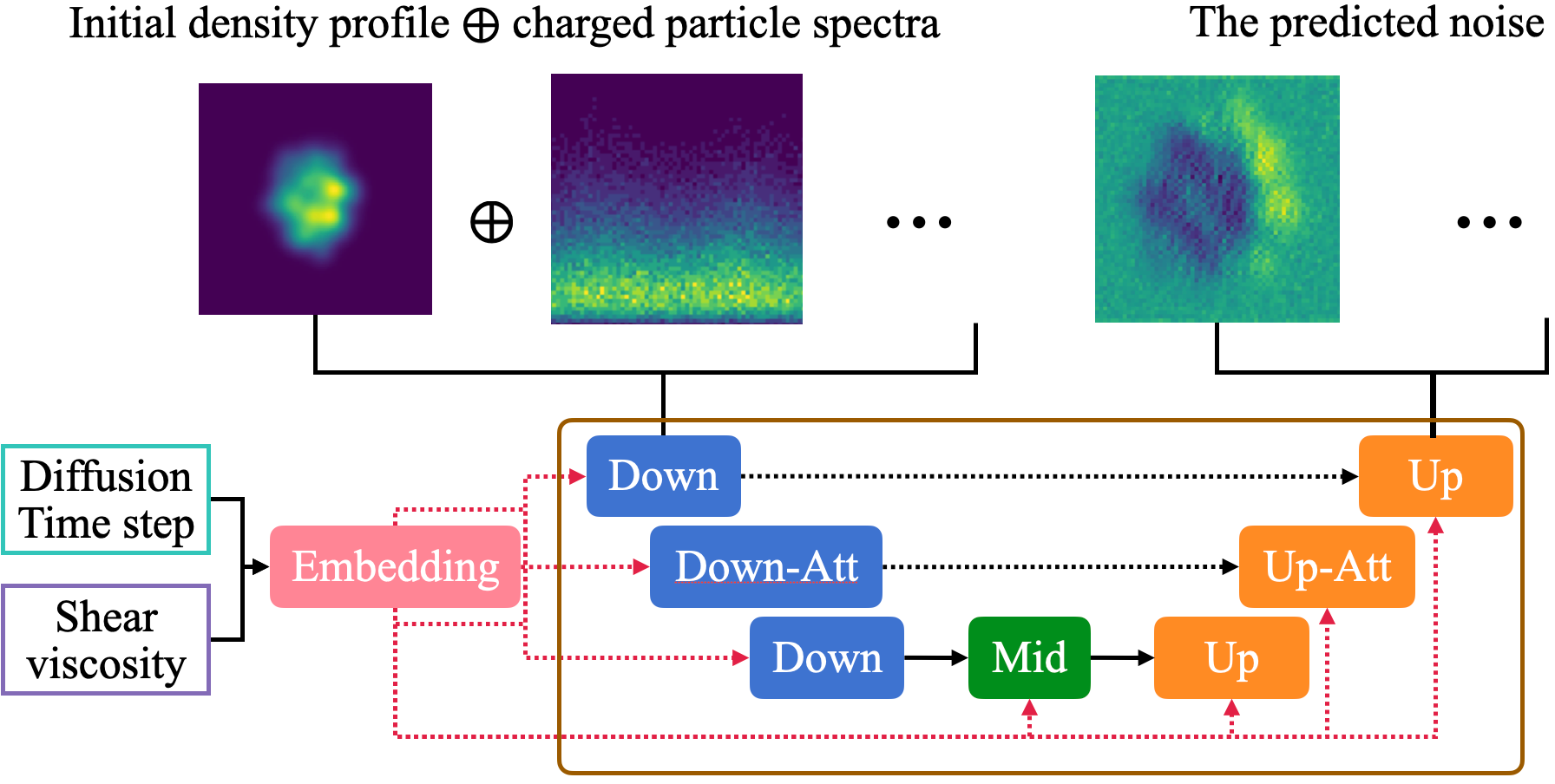}
    \caption{The workflow of DiffHIC. The brown block is the core noise prediction network, which is a typical U-net architecture. The blue and orange boxes are the up and down sampling blocks, respectively. Both are realized by ResNet~\cite{Resnet}. The boxes with "Att" represent the block performing the spatial attention computations~\cite{attention}. {More details of the architecture are given in the Supplemental Material~\cite{NetArchitecture}.}
    The green box is the bottleneck block.  
    }
    \label{fig:diffHIC}
\end{figure}

The conditional reverse diffusion process $p(\pmb S_0|\pmb I,\eta/s)$ is accomplished by learning the scaled score function through a noise-prediction network $\pmb\varepsilon_{\pmb\theta}(\pmb S_t,\pmb I,\eta/s,t)$  as depicted in the workflow shown in Figure \ref{fig:diffHIC}. The initial density profiles and corresponding particle spectra are concatenated channel-wise. The diffusion time steps \(t\) and \(\eta/s\) are encoded via a time-embedding and label-embedding layer, respectively, which are further added together. The input charged particle spectra are noised according to \(\pmb{S}_t = \sqrt{\alpha_t} \pmb{S}_0 + \sqrt{1-\alpha_t} \boldsymbol{\varepsilon}\). A noise-prediction network \(\boldsymbol{\varepsilon}_{\pmb\theta}(\pmb{S}_t, \pmb{I}, \eta/s, t)\) are used, aiming to minimize the mean squared error \(\|\boldsymbol{\varepsilon} - \boldsymbol{\varepsilon}_{\pmb\theta}\|^2\). The training algorithm is summarized in Algorithm 1. Once we have a such trained scaled score function $\pmb \varepsilon_{\pmb \theta} = -\sigma_t \pmb s_{\pmb \theta}$, the particle spectra can be generated from Gaussian noise by solving the reverse ODE (Eq.~\ref{eq:reverseODE}). In this model, the total noise steps is $T=4000$ and we chose a linear noise schedule from $\beta_1 = 0.5\times 10^{-4}$ to $\beta_T = 0.01$. 

\begin{figure}[!t]
    \centering
    \includegraphics[width=1.0\linewidth]{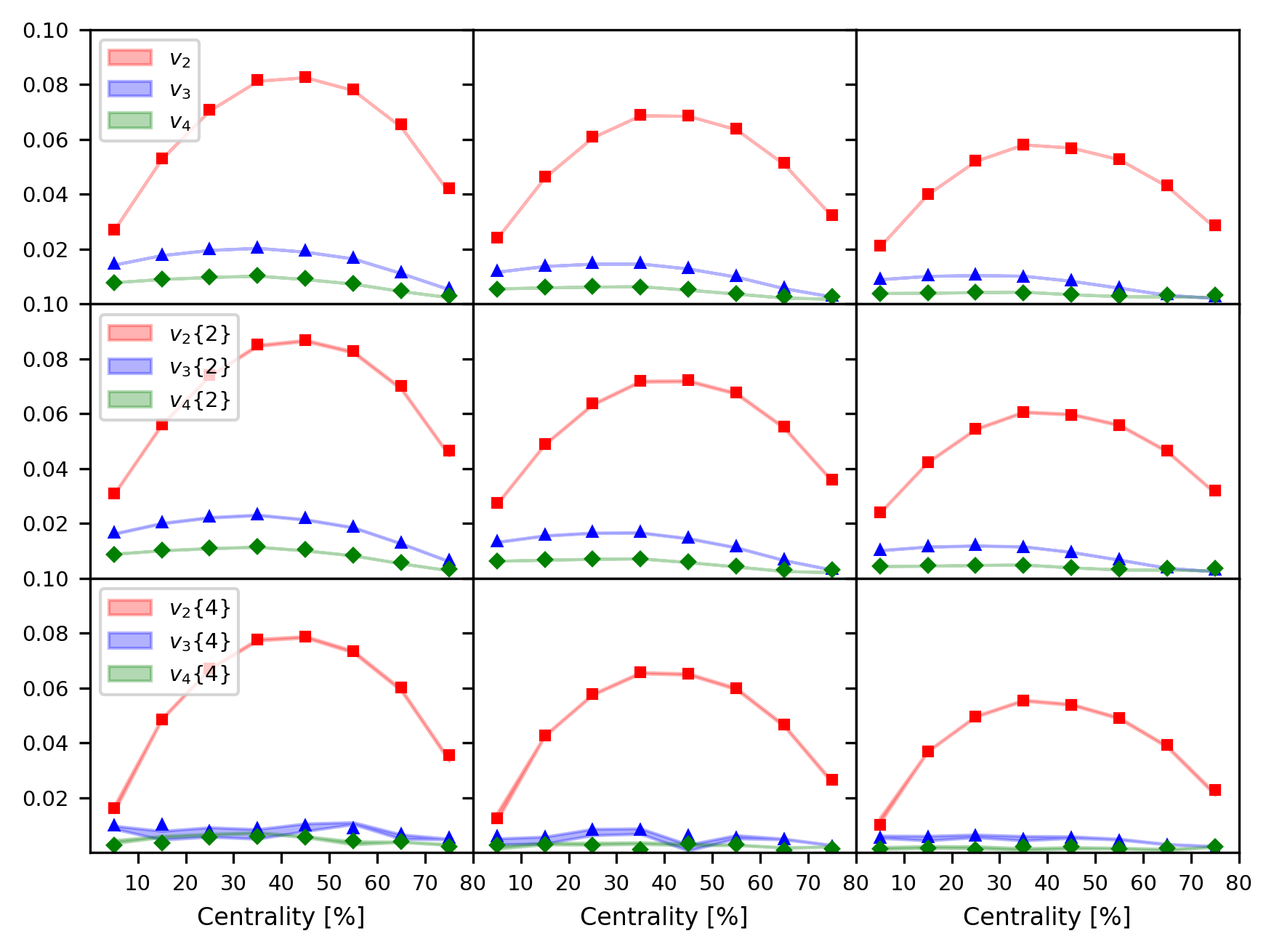}
    \caption{The centrality dependence of integrated anisotropy flow 
    of the 
    {2nd ($v_2$), 3rd ($v_3$) and the 4th ($v_4$) order. }The filled symbols are the ground truth. The first column is the ideal hydrodynamic results. The second and third columns present the results with $\eta/s=0.1,\eta/s=0.2$, respectively. }
    \label{fig:int_flow_cen}
\end{figure}

\begin{figure*}[!t]
    \centering
    \includegraphics[width=1.0\linewidth]{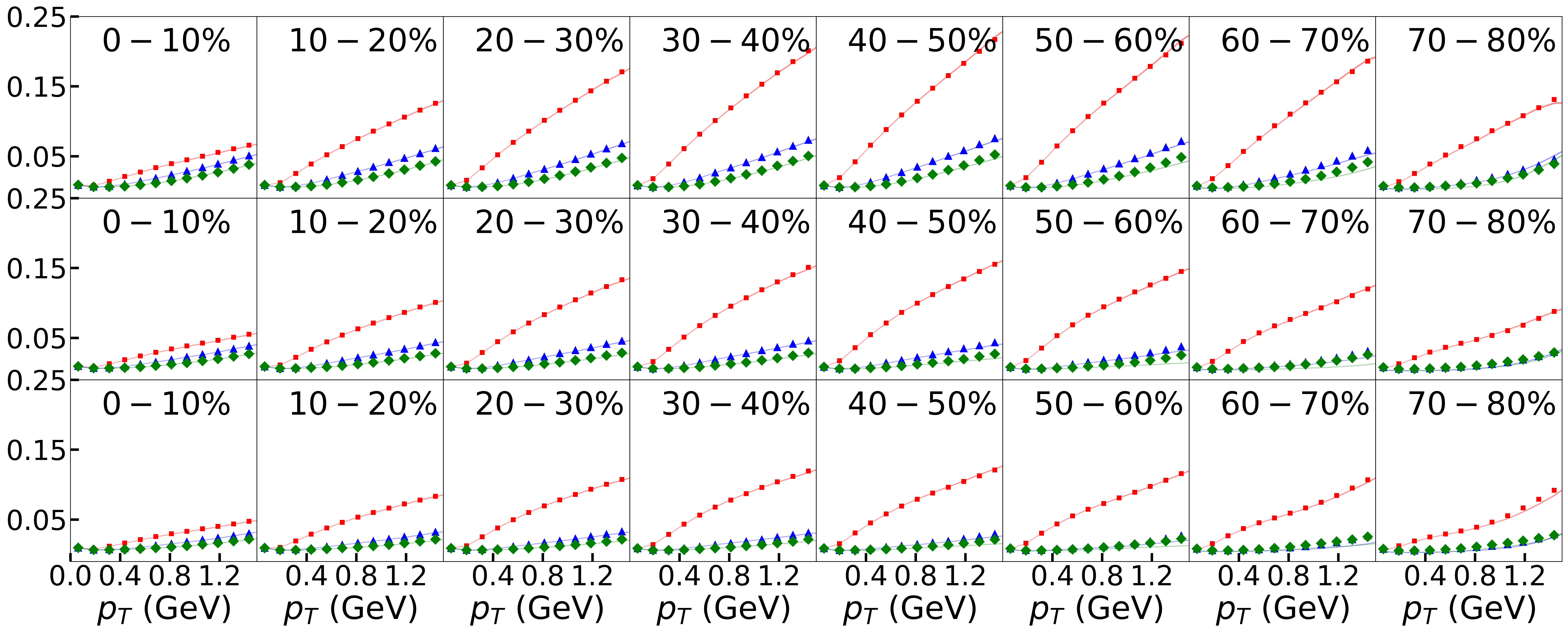}
    \caption{The $p_T$ dependence of anisotropy flow, across all centralities. The filled symbols are the ground truth. The first row is the ideal hydrodynamic results. The second and third rows present the results with $\eta/s=0.1,\eta/s=0.2$, respectively. In each plot, the red, blue, and green lines represent flow of {\color{red}the 2nd order ($v_2(p_T))$, the 3rd order ($v_3(p_T)$) and the 4th order ($v_4(p_T)$)}, respectively.}
    \label{fig:diff_flow}
\end{figure*}

\begin{figure*}
    \centering
    \includegraphics[width=1.0\linewidth]{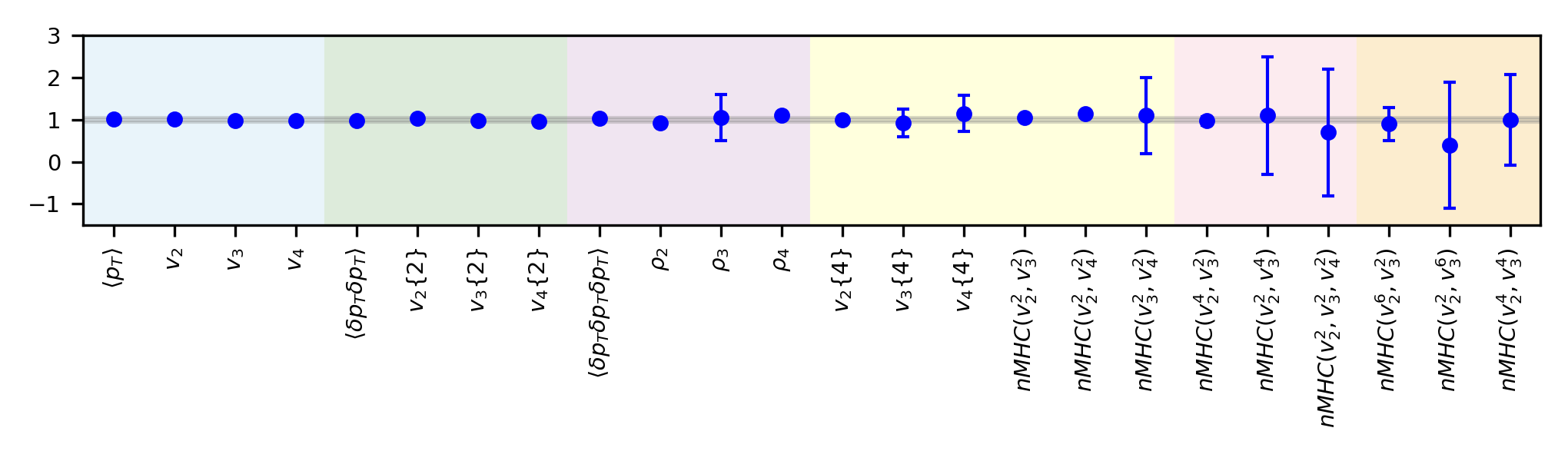}
     \caption{The ratio between the generated results and ground truth in central events. The gray band is $1\pm 0.05$. From left to right, the different color {regions} correspond to flow from single-, 2-, 3-, 4-, 6-, and 8-particle correlations, respectively.
    Errors are estimated via the bootstrap method.
    }
    \label{fig:ratio}
\end{figure*}

\sect{Model Performance}\label{sec:results}
To evaluate the performance of our trained DiffHIC model, we conducted additional 10,000-event simulations for each value of $\eta/s$ 
{in every centrality.} 
We assess the efficacy of DiffHIC by comparing its outputs with 
{the original} numerical simulations, which were regarded as the ground truth.
In terms of the generated particle spectrum, which in practice is a two-dimensional discretized distribution of pixels of size $64\times 64$ (in $p_T$ and $\phi_p$), perfect comparison between DIffHIC and the ground truth is realized with impurities appear only randomly at a level of several pixels.

Below we primarily focus on the anisotropic flow, as they are measurables that quantitatively characterize the spectrum. Anisotropic flow are defined as the Fourier coefficients of the particle spectrum,
\[
{\pmb S}\equiv \frac{dN}{{d^2 {\bf p}_T}} \sim \sum_{n=0} v_n(p_T) e^{i n (\phi_p-\Psi_n)} \,,
\]
with $\Psi_n$ the event plane angle. These flow signatures depend on harmonic order $n$, as well as transverse momentum. Note that the mean transverse momentum, $\langle p_T\rangle$, can be deduced from $v_0$.

Comparisons involving other observables, including in particular correlations and fluctuations among these flow observables~\cite{Shuryak:2014zxa,Heinz:2013th,Alver:2010gr,Ollitrault:1992bk,Romatschke:2009im,flowTeaney:2010vd,flowTeaney:2012ke,flowYan:2013laa},
are detailed in the Supplemental Material~\cite{MoreComparison}. We first present here the integrated flow and differential flow. 
Owing to event-by-event fluctuations, anisotropic flow can be measured with respect to an event plane, or from multi-particle cumulants. In Fig.~\ref{fig:int_flow_cen}, the integrated flow of order $n=2,3,4$ are shown for all centralities, from the event-plane method (top panels), and two-particle cumulant (middle panels) and four-particle cumulant (bottom panels). Comparing to the ground truth (filled symbols), DiffHIC (colored bands) reproduces in general the predictions of the integrated flow, with slight deviation visible for $n=3$ in the four-particle cumulant. The deviation gets reduced as $\eta/s$ increases.
Although differential flow signatures are normally more demanding for model characterizations, in Fig.~\ref{fig:diff_flow}, one observes a perfect comparison between the ground truth (filled symbols) and the DiffHIC predictions (colored bands) for the $p_T$-dependent anisotropic flow, through all centralities. 


To explicitly demonstrate the performance of DiffHIC, the ratios of observables, comparing the generative model to traditional simulations, are shown in Fig.~\ref{fig:ratio}. From left to right, the number of particle correlations involved increases, as indicated by the different color {regions}. These ratios are close to unity, indicating the validity of DiffHIC. Although the model precision decreases systematically as the number of correlated particles increases, it is to some extent anticipated since our simulations are limited by the spectrum resolution of \(64 \times 64\)\footnote{The choice of a 64$\times$64 grid resolution was driven by a careful balance between computational efficiency (GPU memory constraints) and physical fidelity (sufficient to accurately characterize particle spectra). While this resolution provides a robust foundation for our current analyses, we expect that finer resolutions could potentially yield improved precision, especially for higher-order correlations, because it captures finer details of the particle distributions.}
. A higher-resolution model would capture more details of the heavy-ion dynamics. Similar issues also exist in the high-$p_T$ region and peripheral collisions, where effective pixels of the spectrum reduce significantly. Another limitation is the model may struggle in other systems with different collision energies, especially for high-order fluctuations and correlations. Nonetheless, improvement should be achieved by training a model with better resolution. Given the efficiency and flexibility of DiffHIC, fine-tuning our trained model in the new dataset or retraining a specific model is also a possible solution.

\sect{Conclusions and Outlook}
In this {Letter}, the state-of-art generative model (DiffHIC) is for the first time trained to generate the final particle spectra, from event-by-event initial entropy density profiles. While being capable of capturing the two-dimensional distribution in the momentum space accurately, as an end-to-end model, DiffHIC speeds up the numerical simulations by a factor of roughly $10^5$, compared to the traditional approach. Consequently, DiffHIC
alleviates time and resource concerns.

The DifHIC model aims to be applicable to high-precision experimental measurements, establishing a solid theoretical foundation for interpreting observables based on a huge amount of collision data, and physical parameters.
{Notably, DiffHIC is designed to be self-adaptive, enabling the extension of its current parameter set and parameter phase space to incorporate fine structures observed in realistic systems, such as nuclear deformation~\cite{Sun:2025nextDifHIC}. }

\sect{Acknowledgements} Jing-An Sun thanks Xiangyu Wu, Lipei Du, and Han Gao for helpful discussions. {This work is funded in part by the Natural Sciences and Engineering Research Council of  Canada (NSERC) [SAPIN-2020-00048 and SAPIN-2024-00026], in part by the China Scholarship Council, and in part by the National Natural Science Foundation of China under Grants No. 12375133 and 12347106.}
The dataset preparations are made on the Beluga supercomputer system at McGill University, managed by Calcul Quebec and the Digital Research Alliance of Canada.
{The operation of this supercomputer is funded by the Canada Foundation for Innovation
(CFI), Minist\`{e}re de l'\'{E}conomie, des Sciences et de l'Innovation du Qu\'{e}bec
(MESI) and le Fonds de recherche du Qu\'{e}bec - Nature et technologies (FRQ-NT).}

\bibliography{reference}

\end{document}